\documentclass[referee,oldversion,rnote]{aa} % for a referee version
\usepackage{graphicx}
\usepackage{longtable}
\usepackage{txfonts}
\usepackage{rotating}
\usepackage{lscape}
\newcommand{\gsim}{\;\lower.6ex\hbox{$\sim$}\kern-7.75pt\raise.65ex\hbox{$>$}\;}
\newcommand{\lsim}{\;\lower.6ex\hbox{$\sim$}\kern-7.75pt\raise.65ex\hbox{$<$}\;}

\begin{document}
\title{Diluting the material forming the second generation stars in Globular Clusters: the contribution by unevolved stars
 }

\author{
R.G. Gratton\inst{1},
%A. Bragaglia\inst{1}
\and
E. Carretta\inst{2}
%R.G. Gratton\inst{2},
%S. Lucatello\inst{2,3},
%M. Bellazzini\inst{1},
%G. Catanzaro\inst{4},
%F. Leone\inst{5},
%Y. Momany\inst{2,6},
%G. Piotto\inst{7}
%\and
%V. D'Orazi\inst{2}.
}

\authorrunning{R.G. Gratton}
\titlerunning{Less evolved GC stars as source of diluter?}

\offprints{R.G. Gratton, raffaele.gratton@oapd.inaf.it}

\institute{
INAF-Osservatorio Astronomico di Padova, Vicolo dell'Osservatorio 5, I-35122
 Padova, Italy
\and
INAF-Osservatorio Astronomico di Bologna, Via Ranzani 1, I-40127
 Bologna, Italy
%\and
%Excellence Cluster Universe, Technische Universit\"at M\"unchen, 
% Boltzmannstr. 2, D-85748, Garching, Germany 
  }

\date{}

\abstract{In this short communication we consider the possibility that stars 
less evolved than the polluters 
are the source of the dilution needed to explain the observed composition of 
second-generation globular cluster (GC) stars and the Na-O and Mg-Al anticorrelations. If these 
stars can lose 0.5-1\% of their mass during the relevant epochs, there is enough diluting
material to produce the observed anticorrelations. In this case, the original mass 
of proto-GCs was several tens times higher than the current
mass of GCs. While not strictly impossible, this is a stringent hypothesis that needs
more support. Should this
scenario be found true, then the link between the primordial (first-generation) 
population in GC and the field population would be very strong.
  }

\maketitle

\section{Introduction}

For the past few years, it has been known that at least two stellar generations are
present in all globular clusters (GCs) (Gratton et al. 2001; Ramirez \& Cohen 2002; 
see Gratton et al. 2004 for a review). Extensive studies have shown that
the second generation, characterized by overabundance of Na and sometimes
Al, and by a deficiency in O (that has been transformed into N) is dominant, 
since it is responsible for about 2/3 of the stars (Carretta et al. 2009a); the
remaining $\sim 1/3$\ of the stars have a composition that is virtually indistinguishable
from that of field stars. The stars overabundant in Na and deficient in O
are generally distributed over a range of values in [Na/Fe] and [O/Fe], describing an
Na-O anticorrelation and a similar (but not identical) Al-Mg one (Kraft 1994;
see Carretta et al. 2009a and 2009b for several examples). Prantzos \& 
Charbonnel (2006) show that these anticorrelations can be reproduced by 
assuming a typical polluter composition, which is then diluted by different
amounts of pristine material. Such a  dilution
would also help explain the constraints from observed Li abundances, because significant
amounts of this easily destroyed element appear to be present in most 
GC stars (Bonifacio et al. 2002, Pasquini et al. 2005; 
Lind et al. 2009; D'Orazi et al. 2010).

The observed homogeneity in heavy elements (Fe, $\alpha-$ and Fe-group 
species) in GCs and the very peculiar chemical composition of the polluters 
strongly suggest that
they only belong to a limited range of masses, which marginally
contributes to the overall galactic chemical enrichment. When combined with
the large fraction of GC stars belonging to the second generation,
this suggests that only a small fraction of the primordial population remains
bound to the cluster and that the original episode of star formation 
was much larger (Prantzos \& Charbonnel 2006; Carretta
et al. 2010).

While these facts have been assessed well, a number of important issues still 
remain open, precluding full understanding of
the early phases of these massive clusters. They include (i) the nature of the
polluters, which may be either massive asymptotic giant branch (AGB) stars undergoing hot bottom burning
(Ventura et al. 2001) or fast-rotating massive stars (FRMS; Decressin et al. 2007),
and other hypotheses also proposed (e.g. massive binaries, de Mink et al.
2009); (ii) the distribution of stars along the anticorrelation, which might be
either continuous or clumped at some specific value (see e.g. the case of
NGC~2808: D'Antona \& Caloi 2004, Carretta et al. 
2006, etc.); and (iii) the origin of the diluting
material. In this short paper, we focus our attention on this last issue.

D'Ercole et al. (2008) and Decressin et al. (2007) have considered scenarios for
justifying this diluting material. According to Decressin et al. (2007),
second-generation stars form in circumstellar disks around FRMS. 
Depending on the epoch when mass loss occurs, the polluted material
is mixed with various amounts of unprocessed material previously expelled by
the outer regions of the same massive stars, so that stars form with progressively
higher fraction of polluting material. This scenario has serious difficulties
in explaining clumpy distribution of stars along the Na-O anticorrelation,
such as those observed in NGC~2808. Furthermore, it is not clear how this
scenario can generate a bound cluster. 

On the other hand, D'Ercole et al. (2008) proposes that the slow winds from 
the massive AGB stars generate a cooling flow, creating a very dense cloud
from which second-generation stars form. In their scenario, dilution is
obtained by assuming that part of the original gas is left in a torus, which
for some not entirely clear reasons is not polluted by the ejecta of the
core collapse supernovae of the primordial population, and then after some
time falls back onto the dense cloud that is forming the second-generation stars,
hence stars with a progressively larger fraction of diluting
material form. Both a clumpy distribution (see also 
D'Ercole et al. 2010) and a strongly bound cluster can be easily obtained.
However, the hypothesis that a reservoir of pristine gas exists 
appears ad hoc, and somewhat unacceptable.

In this paper we explore an alternative origin for the reservoir of pristine 
gas, which combines concepts from both the Decressin et al. and D'Ercole et al. scenarios. 
We consider the possibility that the reservoir of pristine gas is produced by 
mass lost from less evolved stars (typically - but not only - main sequence stars) than those 
producing the polluting material. This hypothesis has several interesting features:
(i) the mechanism is not ad hoc, since stars of lower mass than the
polluters are surely losing some mass simultaneously to the more massive ones, which are
in later evolutionary phases (actually, this mass should be incorporated
into the hydrodynamical and chemical models); (ii) the material lost by these stars will still be 
unprocessed, but see discussion at the end of Sect. 4; 
(iii) we should not worry about the exact timing of the fallback of 
an unprocessed cloud; (iv) we do not expect this material to be significantly 
enriched by the core collapse supernovae of the primordial population. However, 
this hypothesis may work only if enough mass is lost by the lower mass stars
at the relevant epoch, as discussed in the rest of this paper.

\section{The basic idea}

We need some material with original composition to dilute the polluted one, in
order to (i) reproduce the observed run of O and Na, and (ii) to have some Li saved.
Evidence for this last point comes from NGC~6397, where Bonifacio et al. (2002) find
$\log n$(Li)$\sim 2.3$ in most stars, and from NGC~6752 where Pasquini et al.
(2005) find in Na-rich/O-depleted stars $\log n$(Li)$\sim 1.93$ for
[O/Fe]$=-0.2$~dex. In this hypothesis, Li should be depleted as O.

We note that the dilution factor in GCs is in the range 0-0.7
\footnote{Stars with a dilution $\sim 1$\ can essentially be identified 
with the primordial population.}: 
this is the amount required to place on the Na-O anticorrelation the 
intermediate I stars, that represent the bulk of the cluster population 
(see Gratton et al. 2010; Carretta et al. 2009a).
This means that the amount of primordial, diluting
material is roughly equal to half the amount of processed material.
However, to be conservative, we assumed that the diluting mass is
similar to the polluting one throughout this paper.

How it is possible to achieve this? The explanation 
proposed by D'Ercole et al. (2008) is some primordial gas saved from 
pollution by core-collapse SNe.
An alternative solution, explored in this note, is to consider the
mass lost from the stars less evolved than the polluters.

Our working hypotheses are:
\begin{enumerate}
\item The stars distribute by number as $n(M)=M^{x}$. We assume a lower mass
limit of $0.2~M_\odot$, and an upper limit of $100~M_\odot$. As an alternative,
we use the Kroupa (2001) mass function (between 0.01 and 100~$M_\odot$).
\item The polluted material comes from stars with masses in the mass range 
$M_{\rm min}$ to $M=8~M_\odot$, and they lose all the mass except for 
that is left in compact remnants. This is assumed to be a linear function of
the original mass of the star, which is $0.54~M_\odot$\ for a star of $0.9~M_\odot$,
and $1.24~M_\odot$\ for a star of $8~M_\odot$ (see Ferrario et al. 2005). 
\item The diluting primordial material comes from mass lost by all stars with masses 
less than $M_{\rm min}$ (including low-mass stars).
\end{enumerate}

\section{The toy model}

In our toy model we consider that all stars with masses 
less than $M_{\rm min}$ lose a
fraction $\alpha$ of their mass during the relevant phase. 
For this case, the exponent of the mass function $x$, the fraction of mass 
enclosed in stars in the different mass ranges, the values of
$\alpha$ required to produce as much diluting as polluting mass, and the ratio 
between the mass in the primordial generation and in the GC
\footnote{This is the mass of the GC after formation of the second
generation. Current mass should be lower.}, are listed in 
Table~\ref{t:optiona}.

\begin{table}
\centering
\caption[]{Fraction of mass ($\alpha$) lost by less evolved stars assuming a maximum
polluter mass of $8~M_\odot$}
\begin{tabular}{ccccc}
\hline
 \multicolumn{5}{c}{$M_{\rm min}=4~M_\odot$} \\
\hline
  $x$& Mass fraction    & Mass fraction   & $\alpha$ & Original ratio \\
     & in stars of      & in stars of     &          & 1st gen./GC    \\
     & 0.2-4 $M_\odot$  & of 4-8 $M_\odot$&          &                \\
\hline
-2.9   &     0.948        &	0.025      &0.0208    &    16.9 \\
-2.6   &     0.875        &	0.050      &0.045     &     8.5 \\
-2.3   &     0.727        &	0.083	   &0.090	  &  	5.1 \\
-2.0   &     0.502        &	0.108      &0.170	  &	    3.9 \\
Kroupa &     0.845        & 0.0053     &  0.00489 &     7.0  \\
\hline
 \multicolumn{5}{c}{$M_{\rm min}=5~M_\odot$} \\
\hline
  $x$& Mass fraction    & Mass fraction   & $\alpha$ & Original ratio \\
     & in stars of      & in stars of     &          & 1st gen./GC    \\
     & 0.2-5 $M_\odot$  & of 5-8 $M_\odot$&          &                \\
\hline
-2.9   &     0.959        &	0.0146     &0.0126    &    27.6 \\
-2.6   &     0.895        &	0.030      &0.0279    &    13.4 \\
-2.3   &     0.758        &	0.052	   &0.057	  &  	7.7 \\
-2.0   &     0.540        &	0.070      &0.108	  &	    5.7 \\
Kroupa &     0.864        & 0.035      &0.0331    &    10.9 \\
\hline
 \multicolumn{5}{c}{$M_{\rm min}=6~M_\odot$} \\
\hline
  $x$& Mass fraction    & Mass fraction   & $\alpha$ & Original ratio \\
     & in stars of      & in stars of     &          & 1st gen./GC    \\
     & 0.2-6 $M_\odot$  & of 6-8 $M_\odot$&          &                \\
\hline
-2.9   &     0.965        &	0.0086	   &0.0070     &   49.4 \\
-2.6   &     0.912        &	0.0182	   &0.0158     &   23.3 \\
-2.3   &     0.778        &	0.032      &0.0328     &   13.1 \\
-2.0   &     0.565        &	0.045      &0.062	   &	9.5 \\
Kroupa &     0.878        & 0.021      &0.0181     &   19.8  \\
\hline
 \multicolumn{5}{c}{$M_{\rm min}=7~M_\odot$} \\
\hline
  $x$& Mass fraction    & Mass fraction   & $\alpha$ & Original ratio \\
     & in stars of      & in stars of     &          & 1st gen./GC    \\
     & 0.2-7 $M_\odot$  & of 7-8 $M_\odot$&          &                \\
\hline
-2.9   &     0.971	      &	0.0037	   &  0.0030   & 117.1  \\
-2.6   &     0.917	      &	0.0080	   &  0.0065   &  55.6  \\
-2.3   &     0.736	      &	0.0147	   &  0.0138   &  30.3  \\
-2.0   &     0.589	      &	0.0207 	   &  0.0262   &  21.5  \\
Kroupa &     0.889        & 0.0102     &  0.00715  &  48.8  \\
\hline
\end{tabular}
\label{t:optiona}
\end{table}

The minimum mass required to produce the right amount of unpolluted material
from stars is given roughly by the relation:
$$ M_{\rm min} \sim 15.8 +3.7 \cdot x~~~~~M_\odot,$$
if $\alpha=0.01$. Assuming $x$=-2.3,
the original ratio primordial-to-GC 
is 40.6 and $M_{\rm min}=7.25$; for 
$x$=-2.6 and $x$=-2.9, the minimum masses are $M_{\rm min}=6.65$ and $M_{\rm min}=5.45$, and the
ratios are 36.6 and 34.4, respectively.

An exam of the values listed in Table~\ref{t:optiona} shows that the ratio between the 
original mass of the primordial population and that of the GC is roughly
given by $\sim -0.97/(x\,\alpha)$. Furthermore,
to have the right amount of diluting material, $\alpha$\ should be 
about proportional to the range of masses of the polluters: a wider range
of masses requires a higher value of $\alpha$.

As an alternative case we may consider a given value of e.g. $x=-2.3$\
(or the Kroupa mass function) and assume that 
the range of mass of the polluters is
always equal to $1~M_\odot$. In this way we get the values of Table~\ref{t:table2}.
On the other hand, $\alpha$ represents the total mass lost during the
relevant period. The length of this period depends on the range of mass under
consideration. The mass loss rates required to produce the same value
of $\alpha$ should then scale inversely proportional to this length.

\begin{table}
\centering
\caption[]{Fraction of mass ($\alpha$) lost by less evolved stars assuming various
ranges of polluter masses and either $x=-2.3$ or the Kroupa (2001) mass function) }
\begin{tabular}{ccccc}
\hline
Polluters  &\multicolumn{2}{c}{Mass fraction} & $\alpha$ & Original   \\
Mass Range & in stars of               & in polluters  &          & ratio      \\
($M_\odot$)& 0.2-$M_{\rm min}~M_\odot$ & mass range    &          & 1st gen/GC \\
\hline
\multicolumn{5}{c}{$x=-2.3$}\\
\hline
7-8   &     0.795        &	0.0147	   &0.0138  &   30.3 \\
6-7   &     0.778        &	0.0169     &0.0169  &   25.4 \\
5-6   &     0.756        &	0.0214     &0.0214  &	20.7 \\
4-5   &     0.727        &  0.0288     &0.0280  &   16.3 \\
\hline
\multicolumn{5}{c}{Kroupa mass function}\\
\hline
7-8   &     0.889        &	0.0102	   &0.00715 &   48.8 \\
6-7   &     0.878        &	0.0123     &0.0109  &   36.2 \\
5-6   &     0.864        &	0.0153     &0.0111  &	32.6 \\
4-5   &     0.845        &  0.0199     &0.0150  &   24.9 \\
\hline
\end{tabular}
\label{t:table2}
\end{table}

\begin{table}
\centering
\caption[]{Age, turn-off mass, $M_{TO}$, and mass of AGB stars, $M_{AGB}$}
\begin{tabular}{ccc}
\hline
  Age  & $M_{TO}$ & $M_{AGB}$ \\
\hline
$3.85\,10^7$  & 7.70   & 8.00 \\
$6.3\,10^7$   & 6.07   & 6.34 \\
$1.0\,10^8$   & 4.85   & 5.10 \\
$2.0\,10^8$   & 3.47   & 3.68 \\
\hline
\end{tabular}
\label{t:age}
\end{table}

We quantified the relevant data in Table~\ref{t:age}, using the isochrones 
by Bertelli et al. (2008). The time required for the
evolution of 5-8~$M_\odot$ stars is from $3.85\,10^7$ to $1.1\,10^8$~yr after the 
start of the formation of primordial
generation stars, that is, the mass lost by less evolved stars over an interval of about 
70 Myr may be used to dilute the polluting material. On the other hand,
if the polluters were limited to the range of masses from 7.3 to 8~$M_\odot$,
the interval of time would be only 8~Myr. The same value of $\alpha$ would
then require almost ten times higher mass loss rates. Since the length of the
useful interval of time scales with $M_{\rm min}$\ more rapidly than 
$\alpha$, a higher mass loss rate is required to produce the required dilution if
the range of mass is narrower. Alternatively, if the mass loss rate (for stars
of the same mass) is the same for all GCs, we should expect less dilution
(hence a more clumpy distribution of abundances along the anticorrelation)
for those GCs with a more restricted mass range for polluters.

\section{Limits on $\alpha$}

There are upper limits to $\alpha$\ from other observations and considerations; 
for example, $\alpha$\ for solar type stars is limited because too much mass loss would not be 
compatible with both Li depletion in the pre-main sequence phase and 
observation of the Spite plateau, as well as with the dilution observed in 
normal Population II stars at the base of the subgiant branch.

Unfortunately, the mass loss rates are largely uncertain. For solar type stars, Wood et al. (2002) 
suggest $\dot{M}\sim 4\,10^5/t^{-2.00\pm 0.52}$, but also a saturation at young ages 
(at $2.5\,10^{-11}~M_\odot$/yr), with an uncertainty of a factor 5. The best estimate of the mass
lost during the epoch 40-110 Myr is $2.0\, 10^{-3}~M_\odot$, but the upper
limit is $10^{-2} M_\odot$.
According to Holzwarth and Jardine (2007), a value 100 times solar 
($2\,10^{-14}~M_\odot$/yr) is reasonable for solar type stars in this age range.
This makes the total mass lost of $2\,10^{-12}\times 7\,10^7 \sim 1.4\,10^{-4}~M_\odot$ 
two orders of magnitudes too low over the relevant period.
On the other hand, empirical evidence (Lednicka \& Stepien 2008) suggests an upper
limit of 0.05 $M_\odot$ for Praesepe (0.7 Gyr). If this is lost at a constant
rate, it would give $5\,10^{-2}/7\,10^8$, which is $7\,10^{-11}~M_\odot$/yr.
Integrated over $7\,10^7$~yr, this gives $5\,10^{-3}~M_\odot$.

In turn, the Li-dip may be explained if 0.05~$M_\odot$ is lost from 1.3~$M_\odot$
(solar metallicity) stars (Schramm et al. 1990; see also Russell 1995). The 
proposed mechanism is related to the main instability strip, with a mass loss rate of
$10^{-11} M_\odot$/yr. Integrated over 70~Myr, this yields 
$7\,10^{-4} M_\odot$. 

Summarizing, an upper limit to mass lost during the 40-110~Myr epoch for a solar
type star is 0.01~$M_\odot$. This implies a maximum mass loss rate of 
$\sim 10^{-10}~M_\odot$/yr. On the other hand, in our approach, $\alpha$ is
a suitable average value for the fraction of mass lost during this phase
by {\it all stars that are less evolved than those causing the pollution}.
This average also includes (i) low-mass stars that are still in the 
pre-main sequence phase at the relevent epoch ($M<0.5~M_\odot$: 
Di Criscienzo et al. 2009)\footnote{We notice that T Tau stars 
have disks with masses of $\sim 0.01$~M$_\odot$, which however are probably 
dissipated on a timescale of $\sim 6$~Myr (Hillenbrand et al. 1998; Haisch,
Lada \& Lada 2001). The dissipation time might be longer for stars of
lower mass and brown dwarfs (Carpenter et al. 2006; Apai, Luhman \& Liu, 
2007).}; (ii) the same stars later providing the 
polluting material, but in earlier evolutionary phases (e.g. while they are
in the Cepheids instability strip, see e.g. Neilson \& Lester 2009); and 
(iii) interacting binaries (see e.g. Mennickent et al. 2010). All these classes
of objects may have mass-loss rates (in units of the stellar mass) that are significantly
more than solar type stars of similar ages. The assumption of a value
$\alpha \sim 1$\% therefore does not appear completely implausible.

It should, however, be noticed that the chemical composition of the winds from these
other classes of stars might not precisely reflect the original composition.
Low-mass stars should have burnt Li during the pre-main sequence phase,
when they were still fully convective. If they contributed significantly
to the dilution required to explain the O-Na anticorrelation, then the Li 
observed in second-generation stars with a high degree of dilution should have been 
produced by the same polluters, and there should not be a simple Na-Li
anticorrelation. As a matter of fact, this is not at all excluded by
current observations (see e.g. D'Orazi \& Marino 2010). For
the other classes of objects, Luck \& Lambert (1992) made a chemical abundance 
analysis of Cepheids in the LMC and SMC. The [O/Fe], [Na/Fe], [Mg/Fe], and 
[Al/Fe] ratios are roughly solar, so that a putative wind from these stars 
may contribute to the dilution needed to explain the Na-O and Mg-Al 
anticorrelations. However, these stars are depleted in C and Li, with 
evidence of dredge-up of N-rich material. For interacting 
binaries, depletion of C and excesses of N have been found in most Algols 
(Parthasarathy et al. 1983; Tomkin et al. 1993), and
even O is depleted in the more massive $\beta$~Lyr (Balachandran et al. 1986;
indeed, massive binaries have been proposed as a source of the polluting 
material, rather than as diluters: de Mink et al. 2009). If these two classes
of objects were indeed the source of the diluters, we should then expect
that the Na-O and the C-N anticorrelations should appear quite different. 
This prediction might be compared with the actual data (see e.g. 
Briley et al. 2004).
 
\section{Limits on the ratio primordial population/GC}

The ratio of primordial population/GC cannot be greater than the ratio between
the mass of the halo and the mass in GCs, which is $\sim 100$. In the current
framework, this provides a lower
limit for $\alpha$, that is, $\alpha>-0.0097/x$.

If we assume $\alpha=0.01$\ and $x=-2.3$, we should expect a mass ratio
of primordial population to the GC of $\sim 40$. In this framework,
to produce a cluster with a current mass of $10^6~M_\odot$ (like NGC~2808 or 47~Tuc
\footnote{We neglect here the mass lost by the GC after its formation.}), one 
should start with $\sim 4\,10^7~M_\odot$ of gas with a very homogeneous 
composition ($\Delta$[Fe/H]$<<0.04$~dex); possibly a value of $\sim 8\,10^7 
M_\odot$ for the cloud mass would be more reasonable, considering a 
reasonable $\sim 0.5$ star formation efficiency for the primordial (original) 
generation. Star formation for the primordial generation cannot have lasted very long, or else SNe 
would have either (i) stopped it or (ii) contributed to the nucleosynthesis. A 
reasonable upper limit is $10^7$~yr. Such a massive ($4\,10^7~M_\odot$, stellar mass) 
star-forming region should be very luminous ($M_V \sim -18$).

Almost all (98.7\%) of the original cluster population would have been lost from 
the expansion of the cluster after SNe explosion and mass loss. The second 
burst of star formation (the one producing the second generation) would then 
be a much smaller episode. With a mass of $5\,10^5~M_\odot$ of stars 
produced, it should have a luminosity of $M_V=-14$\ at peak. At this 
epoch, the fading original population should still have a luminosity of $M_V=-
16$, dominating the compact central cluster of second-generation stars.

Assuming a ratio $M/L=2$, all galactic GCs summed up have a mass of 
$3.4\,10^7~M_\odot$.
The primordial population should then be $\sim 1.4\,10^9~M_\odot$. This
requires that roughly half of the halo/thick disk mass comes from GCs
(see Gratton et al. 2010, in preparation, where we will discuss
the metallicity distribution and the element-to-element abundance ratios of the 
primordial population, which are very similar to that of the halo-thick disk).

\section{Can the wind from main sequence stars contribute to the second generation?}

Winds from main sequence stars are quite fast, with typical velocities of
a few hundred km/s (Neugebauer 1994; Dupree 2005). This is much higher than 
the escape velocity from a proto-GC. Could this wind be contributing to the
second generation?
To address this issue, we follow the same arguments considered by Smith
(1999), who considered the case for dissipation of intracluster gas (lost at
low speed by red giants) from the energy injection by the high-velocity
wind from the main sequence. We should replace the wind by red giants with the 
much stronger winds of AGB stars. The specific mass-loss rate for massive
AGB stars is about two orders of magnitude more than that for current red 
giants: $\alpha_{\rm AGB}\sim 3\times 10^{\-17}$~s$^{-1}$. On the other hand,
the specific mass loss rate for the diluters should be comparable to that of
the AGB stars, or else dilution would be negligible.

Neglecting cooling for the moment, the winds from the diluters are kept within
the cluster if Eq. (8) of Smith (1999) holds. Assuming 
$\alpha_{\rm MS}/\alpha_{AGB}=0.5$, this equation requires that the
wind velocity from the polluter $v_{\rm MS}$\ satisfies the relation
$v_{\rm MS}<5.2~\sigma_{\rm los}$, where $\sigma_{\rm los}$ is the
line of sight velocity spread of the proto-GCs. Current values of
$\sigma_{\rm los}$ are $\leq 20$~km/s for almost all GCs. It is likely
that $\sigma_{\rm los}$ could have been slightly higher for proto-GCs.
We conclude that winds with velocities $<100$~km/s were almost certainly kept 
within the clusters. These include potential polluters like red giants in those
phases preceding the AGB, close binaries, and perhaps surviving disks around
low-mass stars.

On the other hand, even faster winds could contribute to dilution if
the input kinetic energy is dissipated by radiative cooling. This may
happen if radiative cooling losses are greater than the heating due
to thermalization of the diluters' winds. To estimate when this may occur,
we use inequality (21) of Smith (1999), which must be solved
simultaneously with his Eq. (10). In this inequality, we should consider the
radiative cooling coefficient appropriate to the central temperature
of the intracluster wind, which can be taken from Faulkner \& Freeman (1977),
as done by Smith (1999). Assuming a protocluster (core) radius of 1 pc and
central density of $10^5$~M$_\odot$~pc$^{-3}$ (which are typical values
for present GCs), we find that cooling was efficient at the relevant epoch
insofar $\alpha_{\rm MS}/\alpha_{\rm AGB}\leq 0.2$. This value should 
be taken only as an order of magnitude estimate, owing to the approximations
made and because several quantities are not well known. For instance, it
assumes a wind velocity from MS stars of 600 km/s, and is valid for the
radius and densities considered above. It is possible that
proto-GCs were more compact ($\sim 0.5$~pc) and much denser ($>10^6$~M$_\odot$~pc$^{-3}$)
(Marks \& Kroupa 2010). If we use the values for the original radius and
central density from these authors, we get typical upper limits of 
$\alpha_{\rm MS}/\alpha_{\rm AGB}\leq 0.6$, which is compatible
with a significant dilution by winds from main sequence stars.

We conclude that the wind from stars that are less evolved than the polluters
might have contributed to the diluting material. This is very likely if a 
substantial fraction of this wind is at velocity $<100$~km/s, but it
is still possible even for faster winds, at velocities of a few hundred
km/s.

\section{Conclusions}

Assuming that the fraction of mass lost is  roughly independent of stellar mass, then
mass loss during MS (or more in general during the relevant period by stars
less evolved than the polluters) may provide the diluting mass only if $\sim 1$\% (within a
factor of 2) of the mass is lost in the relevant phase and the ratio of masses
between the primordial generation and the GC is greater than 40 (still within a factor of 2). 
This  requires that a
considerable fraction of the halo (at least half) originated in  proto-GCs, 
and also that these proto-GCs lost about $>95$\% of their primordial
generation stars.
Although we do not have direct access to events that occurred a Hubble time
ago, chemical abundances allow us to put strong constraints even on those
early phases of GC evolution. It is currently well known that second-generation 
stars are formed from the ejecta of massive stars of the primordial
generation (e.g. Gratton et al. 2001). From our ongoing FLAMES survey we also
know the current proportions of primordial and second-generation stars in GCs. The
latter represent the bulk of stars in GCs (about 2/3) whereas the primordial
component is still present, but at a level of only a third of the current
stellar population. Taken together, these two observations mean that a large
fraction of stars of the primordial generation must necessarily be lost from GCs.

Theoretical considerations about star formation efficiency (e.g. Parmentier et
al. 2008), violent relaxation (Lynden-Bell 1967) and gas expulsion (e.g.
Baumgardt et al. 2008) predict that a huge mass loss occurs in early phases
of the GC lifetimes. 

In Carretta et al. (2010) we explored the estimates of the initial mass of the
primordial generation in GCs required to satisfy these two
observational features. With suitable (and realistic) assumptions on the initial
mass function of primordial and second-generation stars, mass ranges of the
preferred polluters (either AGBs or FRMS), and initial-final mass relation we
were able to show that a proto-GC should have lost $\sim 90\%$ of its primordial
stellar population. This is a value that is not very distant from the $>95$\% we derive in
the present paper.

As mentioned in Sect. 2, we assumed that the diluting mass is equal to the 
polluting one. Actually, it might be half of this value, so relaxing these 
constraints a bit. We conclude that, on the whole, this scenario is then quite 
constraining and not entirely implausible. Of course, a better determination of 
the mass loss from young stars would be highly welcomed.

\begin{acknowledgements}
We thank the referee for her/his very interesting comments, as
well as for reminding us of the very interesting paper by Smith (1999).
We thank Franca D'Antona and Annibale D'Ercole for useful discussions and 
Valentina D'Orazi for a critical reading of the paper. This research
has been funded by PRIN MIUR 20075TP5K9.
\end{acknowledgements}

\end{document}